\documentclass[a4paper,11pt]{iopart}

\usepackage{graphicx}
\usepackage{caption}
\usepackage{subcaption}
\usepackage{dcolumn}
\usepackage{bm}
\usepackage[dvipsnames]{xcolor}
\usepackage{upgreek}
\usepackage{amssymb}
\usepackage{verbatim}
\usepackage{appendix}
\usepackage{listings}
\usepackage{units}
\usepackage{fancyhdr}
\usepackage{csquotes}
\usepackage{hyperref}
\usepackage{gensymb}
\usepackage[normalem]{ulem}

\usepackage{etoolbox}
\makeatletter
\patchcmd{\hyper@makecurrent}{%
    \ifx\Hy@param\Hy@chapterstring
        \let\Hy@param\Hy@chapapp
    \fi
}{%
    \iftoggle{inappendix}{
        \@checkappendixparam{chapter}%
        \@checkappendixparam{section}%
        \@checkappendixparam{subsection}%
        \@checkappendixparam{subsubsection}%
        \@checkappendixparam{paragraph}%
        \@checkappendixparam{subparagraph}%
    }{}%
}{}{\errmessage{failed to patch}}

\newcommand*{\@checkappendixparam}[1]{%
    \def\@checkappendixparamtmp{#1}%
    \ifx\Hy@param\@checkappendixparamtmp
        \let\Hy@param\Hy@appendixstring
    \fi
}
\makeatletter

\newtoggle{inappendix}
\togglefalse{inappendix}

\apptocmd{\appendix}{\toggletrue{inappendix}}{}{\errmessage{failed to patch}}
\apptocmd{\subappendices}{\toggletrue{inappendix}}{}{\errmessage{failed to patch}}

\definecolor{hreflinkcolor}{rgb}{0.13,0.17,0.83}
\hypersetup{colorlinks=true,urlcolor=hreflinkcolor,
		linkcolor=hreflinkcolor,citecolor=hreflinkcolor}

\begin{document}

\title{Generation of attosecond electron bunches and X-ray pulses from few-cycle femtosecond laser pulses}
\author{J.~Ferri$^{1,2}$, V.~Horn\'y$^1$ and T. F\"ul\"op$^1$}
\address{$^1$ Department of Physics, Chalmers University of Technology,
  SE-41296 G\"{o}teborg, Sweden\\
  $^2$ Department of Physics, University of Gothenburg, SE-41296 G\"{o}teborg, Sweden  }
  
  \ead{julien.ferri@physics.gu.se} 
\date{\today}

\begin{abstract}
Laser-plasma electron accelerators can be used to produce high-intensity X-rays, as electrons accelerated in wakefields emit radiation due to betatron oscillations. Such X-ray sources inherit the features of the electron beam; sub-femtosecond electron bunches produce betatron sources of the same duration, which in turn allow  probing matter on ultrashort time scales. In this paper we show, via Particle-in-Cell simulations, that attosecond electron bunches can be obtained using low-energy, ultra-short laser beams both in the self-injection and the controlled injection regimes at low plasma densities. However, only in the controlled regime does the electron injection lead to a stable, isolated attosecond electron bunch. Such ultrashort electron bunches are shown to emit attosecond X-ray bursts with high brilliance.
\end{abstract}


\section{Introduction}
\label{sec:intro}
Laser wakefield acceleration (LWFA) is an alternative technique for accelerating electrons to relativistic energies~\cite{Tajima1979}. In LWFA, an intense laser pulse propagates through a plasma, expelling electrons while leaving the ions almost stationary on the laser timescale. This charge separation drives a plasma wave in the wake of the laser pulse (hence the term wakefield). Such laser-produced plasma wakefields can sustain orders of magnitude higher accelerating fields than conventional accelerators. If electrons are injected at the right location and with the right energy, they will remain in the accelerating phase of the plasma wave electric field, and gain relativistic energies. 
Since the first experiments demonstrating mono-energetic multi-MeV electron beams \cite{Mangles_Nature_2004,Geddes_Nature_2004,Faure_Nature_2004} there has been a rapid growth in the achievable energy and quality of the electron beams. Currently available LWFAs reach up to  GeV-energy levels in only a few centimeters~\cite{wang2013, Gonsalves2019}. Furthermore, increased control over the electron beam parameters has also been achieved, in particular through the development of techniques for electron injection in the plasma wakefield.  One such example is the so-called gradient injection, achieved by introducing a density down-ramp early in the electron acceleration region~\cite{Bulanov1998, Geddes2008}, which allows for the tunability of the injected charge and of its energy profile~\cite{Gonsalves2011, Hansson2015}.

In addition to their energy gain, electrons perform transverse oscillations in the wakefield, which causes the emission of synchrotron-like hard X-rays in a process known as betatron radiation~\cite{Esarey2002, Rousse2004, Corde_RevModPhys_2013}. This X-ray source inherits the features of the electron beam: micrometer source size~\cite{Shah2006}, femtosecond duration, and low divergence that can be in the mrad range~\cite{Kneip2010}, leading to a high-brilliance source. Such characteristics opened up a range of advanced applications of betatron sources, such as high resolution phase contrast imaging~\cite{Kneip2011, Fourmaux2011} or femtosecond X-ray absorption spectroscopy~\cite{Mahieu2018}.

In parallel to the course toward higher-energy electron beams utilizing petawatt laser facilities, attempts are being made for generating electron beams with low-energy laser facilities, which benefit from a high repetition rate (kHz). Monoenergetic beams have been obtained experimentally  in such mJ-class laser systems~\cite{Schmid2009,Guenot2017}. Additionally, theoretical studies have determined the role of laser parameters, such as carrier envelope phase effects (CEP) or frequency chirp, to tune the electron energy~\cite{Lifschitz2012, Beaurepaire2014}.

An important feature of LWFA-generated electron beams is their short duration, which has been characterized to be in the femtosecond range in experiments using coherent radiation transition~\cite{Lundh2011}. Recent theoretical studies proposed different ways to obtain sub-femtosecond duration electron bunches from LWFA, including plasma density tailoring~\cite{Tooley2017}, crossing laser pulses~\cite{Horny2019} or injection suppression with a magnetic field~\cite{Zhao2019}. Attosecond electron beams would be a major achievement, as it would open the door to secondary attosecond X-ray sources, which would allow probing matter on ultrashort time scales. 

In this paper, we present a systematic study of the generation of sub-femtosecond electron bunches produced by low-energy, ultra-short laser beams through quasi-cylindrical Particle-in-Cell (PIC) simulations. We predict that attosecond electron bunches can be obtained both in the self-injection and the controlled injection regime at low plasma densities, but only the controlled regime leads to a stable, isolated attosecond electron bunch. Furthermore, we investigate the radiation emission from these bunches, and show that brilliant attosecond X-ray bursts are generated.

\section{Sub-femtosecond electron bunch generation}
We consider the interaction of an $800~$nm-wavelength laser pulse with a fully-ionized gas-jet.
The laser pulse has a Gaussian profile in space and time and is linearly polarized in the $y$ direction. Its duration is $\tau_0 = 8.3~$fs (FWHM of the intensity), energy $E_0 = 50~$mJ, and it is focused on a $3.8~\upmu$m spot size (FWHM of the intensity) at the entrance of the pre-ionized electron plasma, yielding a normalized vector potential $a_0 = 3.44$. Ions are considered immobile.
Simulations are run with the fully relativistic quasi-cylindrical PIC code CALDER-Circ~\cite{Lifschitz2009}, using 3 Fourier modes for the decomposition of the currents and fields. We use a temporal time step $0.114\omega_0^{-1}$, and longitudinal and radial resolution $0.125c\omega_0^{-1}$ and $0.625c\omega_0^{-1}$ respectively, with $c$ the speed of light and $\omega_0$ the laser frequency. The simulation uses a moving window consisting of $1500\times300$ cells, in which the electrons are represented by 64 macro-particles per cell. Moreover, Maxwell's equations are solved using a scheme that reduces numerical Cherenkov radiation~\cite{Lehe2013}.

\subsection*{Self-injection}
We first consider a gas jet consisting of a $30~\upmu$m-long linear density ramp-up followed by a plateau with a density that is varied between $n_0=2\times 10^{19}$~cm$^{-3}$ and $n_0=8\times 10^{19}$~cm$^{-3}$.

We note that in this optimal regime, massive self-injection occurs and that a quasi-monoenergetic peak is observed at high energies. The energy of this peak is close to the prediction of the scaling laws, which yield an energy of about 51~MeV for a plateau density $n_0=4\times 10^{19}$~cm$^{-3}$, slightly above the value reached in the simulation ($\sim47$~MeV). The monoenergetic component of the electron energy distribution disappears when the plasma density becomes too far from the matched one predicted by the theory. At higher densities the injected charge continues to grow but reaches lower averaged energies. Finally, at lower densities, the injected charge decreases quickly, with no charge being injected when the density is lowered to $n_0=2\times 10^{19}$~cm$^{-3}$.

\begin{table}
    \centering
    \begin{tabular}{c||c|c|c|c|c}
         $n_0$ (10$^{19}$~cm$^{-3}$) &2 &3 & 4 & 5 & 8\\ \hline
         $L_{\mathrm{opt}}$ ($\upmu$m) & 220  & 220  & 170  & 130  & 70 \\ \hline
         $E_{mono}$ (MeV) & - & 60--90 & 47 & 42 & -\\ 
    \end{tabular}
    \caption{Optimum acceleration length in the plateau, $L_{\mathrm{opt}}$, after which the maximum electron energy is reached, and energy of the quasi-monoenergetic peak, $E_{mono}$, as a function of the plasma density. For the lowest and highest densities, the spectra do not exhibit a quasi-monoenergetic peak. 
    }
    \label{tab:tab1}
\end{table}

\begin{figure}[t!]
  \includegraphics[width=1.0\linewidth]{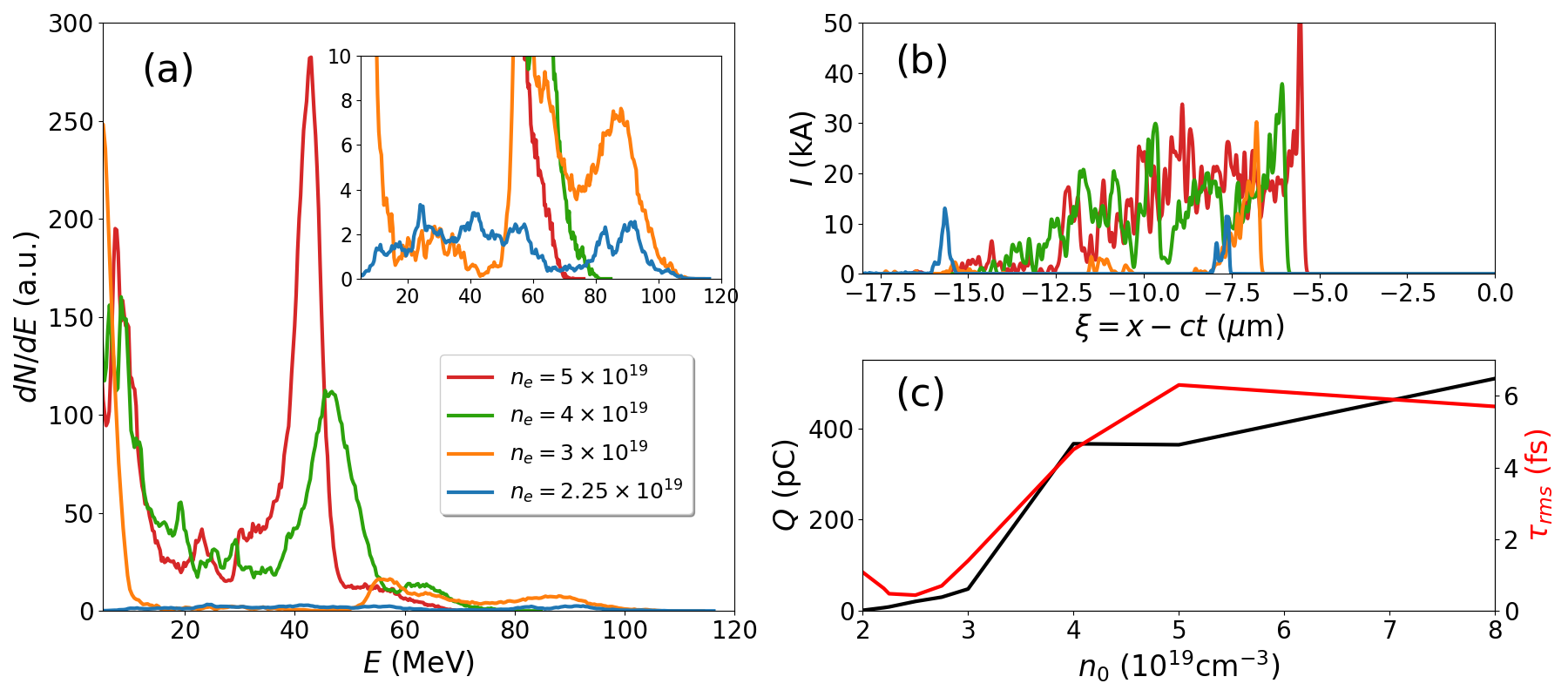}
  \caption{(a) Electron energy spectra after the optimal acceleration length defined in table \ref{tab:tab1} for different densities and (b) current profiles of these electron beams. The longitudinal position is indicated as a function of $\xi=x-ct$, with $\xi=0$ corresponding to the laser position. (c) Injected charge $Q$ (black line) and RMS duration $\tau$ of the injected beam (red line) as a function of the plasma density.}
\label{fig:fig1}
\end{figure}

As the laser pulse used in the simulations is much shorter than the ones typically used in the LWFA context, leading to higher densities in the matched regime, it may be expected that shorter electron beams can be obtained in the interaction conditions ensuing from the scaling laws.  However, we find that the duration of the electron beam is not particularly short at the densities which are optimal for electron acceleration. Current profiles of these electron beams are shown in Fig.~\ref{fig:fig1}(b), and exhibit RMS duration $>4~$fs when the density is above $n_0=4\times 10^{19}$~cm$^{-3}$. This is due to the high plasma densities that lead to massive injection in this regime. For the lower densities, the injection becomes more localized in space as we barely exceed the threshold condition for electron injection, and self-injection in the first and second cavities of the wakefield leads to two well-defined electron bunches. For example at $n_0=2.25\times 10^{19}$~cm$^{-3}$ (blue line), the charge and duration of the first electron beam at about $\xi=x-ct=-8~\upmu$m are $8.5$ pC and $0.48$~fs, respectively. This is not an isolated attosecond beam, however, as a second beam of $11.6$~pC charge is injected in the second cavity about $8~\upmu$m behind the first one. Note, that the spacing between the two peaks of the current decreases as the density rises (visible for densities below $n_0=3\times 10^{19}$~cm$^{-3}$), as it corresponds to a shorter wakefield period with the shorter plasma wavelength.

Figure~\ref{fig:fig1}(c) exhibits the variation of the injected charge $Q$ and duration $\tau$ of the first beam as a function of the injection density. Overall, the charge and duration offer a very similar dependence on the density, leading to long-duration and high-charge electron beams at high densities. For lower densities, of the order of $2.3$--$2.5\times 10^{19}$~cm$^{-3}$, very close to the injection threshold, as the injected charge becomes arbitrary small, a minimum duration is obtained of about half a femtosecond. However, such short duration can only be obtained for a narrow range of parameters. The duration of the injected beam increases again when the density is further decreased. This variation of the duration for small changes of the density illustrate the fact that injection at densities just over the threshold becomes very irregular. It will indeed depend on tiny variations of the potential vector $a_0$ or of the density profile, and will thus very likely exhibit high shot-to-shot variation in an experiment.
In the following, we will investigate how this picture can be changed by controlled injection.

\begin{figure}[t!]
  \includegraphics[width=0.9\linewidth]{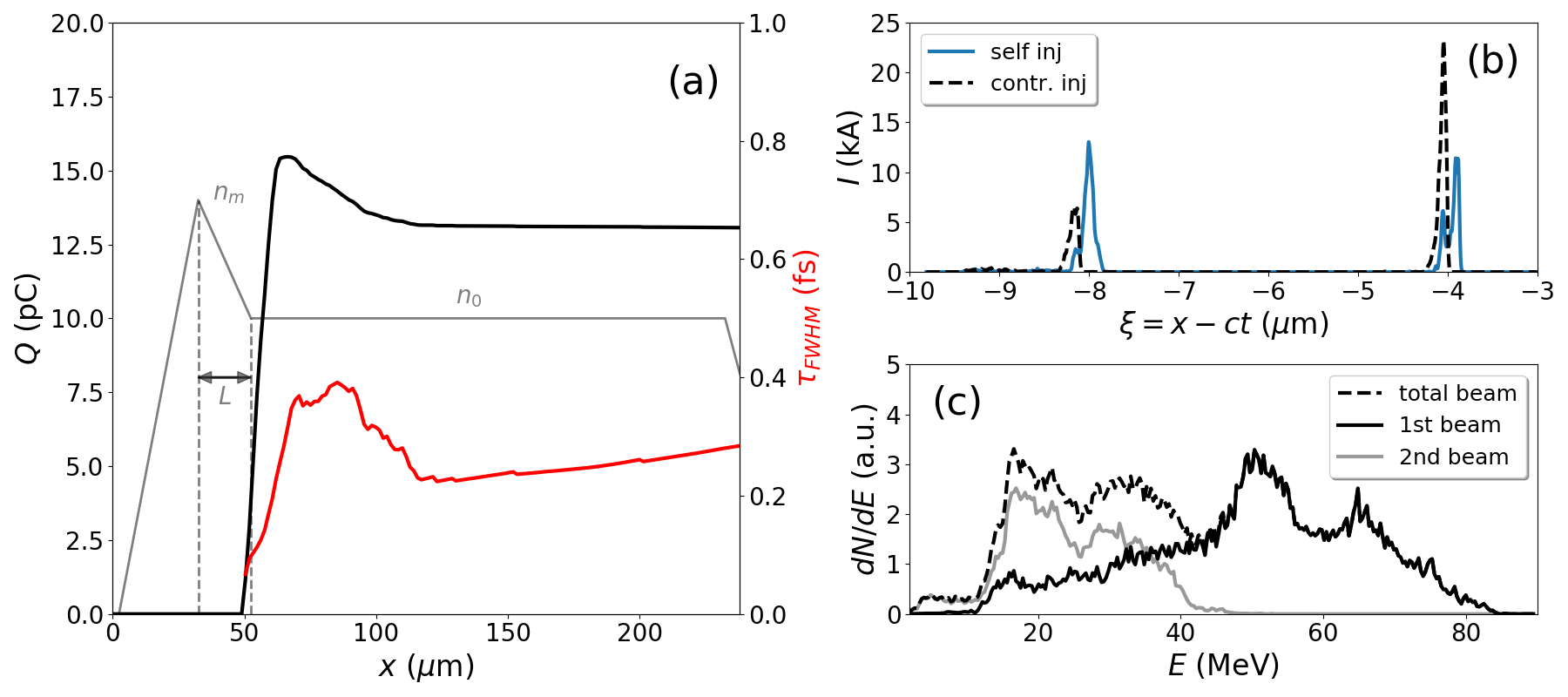}
  \caption{(a) Evolution of the trapped charge $Q$ (black line) and of the temporal duration $\tau$ of the electron bunch (red line) with the propagation length for the case with a density peak and  $n_m$ = 1.04. The density profile used for the simulations with down-ramp injection is indicated in grey lines. (b) Electron beam current after $230~\upmu$m, with $n_0 = 2.25\times 10^{19}$~cm$^{-3}$ without density peak (solid blue) and with $n_m$ = 1.04 and  $n_0 = 2.0\times 10^{19}$~cm$^{-3}$ (dashed black). The longitudinal position is indicated as a function of $\xi=x-ct$, with $\xi=0$ corresponding to the laser position. (c) Energy spectra of the down-ramp injected beam represented in (b), and of the beam injected in the first (plain black) and second cavities (plain grey) only.}
\label{fig:fig2}
\end{figure}

\subsection*{Controlled injection}

In order to improve the control over the beam parameters, we introduce a density down-ramp in the plasma density profile. Thus, we use the same density profile as before, except that we add a high-density peak at the beginning of the propagation (see Fig.~\ref{fig:fig2}(a)). Such sharp density transitions can be implemented by inserting a razor or knife edge into the gas jet~\cite{Schmid2010, Dopp2018} or by designing the gas flow from an array of Laval nozzles. The density of the plateau is lowered to $n_0 =  2\times 10^{19}$~cm$^{-3}$ to avoid unwanted self-injection during the propagation, and the peak of the density $n_m$ is varied from $n_0$ up to $1.4n_0$. The length of the down-ramp is first fixed at $L=10~\upmu$m. Using this density profile, the injection is stabilized with the down-ramp, leading to the stable generation of shorter attosecond electron beams. Short durations are obtained using density peaks with values barely above $n_0$, in agreement with previous results~\cite{Tooley2017}. 

Importantly, isolated attosecond electron beams can be generated, as shown in Fig.~\ref{fig:fig2}(b). Using a density peak characterized by $n_m = 1.04n_0$, a 13.1~pC, 0.28~fs-long (RMS) electron beam can be generated, while the charge in the second beam is reduced to 6.2~pC.
The second beam contains, on average, more charge than the first beam in the self-injection case, while it contains only about 40~\% of it when the injection is controlled by the down-ramp. Furthermore, the beam injected in the second cavity is accelerated to lower energy (see Fig. \ref{fig:fig2}(c)). 

Interestingly, the beam durations that are achieved during the acceleration can be substantially shorter than the final values. The evolution of the charge and duration of the first beam is depicted in Fig.~\ref{fig:fig2}(a). From the increase of the charge (black line) after $x=50~\upmu$m, we can deduce that the injection is actually triggered by the density down-ramp. The trapped charge then decreases slightly as some low-energy electrons at the rear of the bunch are laterally ejected, and remains constant after $x\sim120~\upmu$m. The duration of the beam (red line) follows a similar pattern, and increases with the charge in early times, before decreasing as some electrons from the rear are ejected. It reaches a minimum at $x=120~\upmu$m, slightly above 200~as. However at constant charge, the duration slowly increases during the acceleration, due to the energy spread of the electron beam. The lower energy electrons at about 20~MeV are slowly dragging behind, contributing to an increase of the beam duration. Within a distance of $100~\upmu$m, such electrons indeed accumulate a delay of about 100~as compared with the most energetic electrons in the beam. Note that this intrinsic limitation of the duration can only be overcome by reducing the energy spread of the electron beam, or by reducing the acceleration distance.

\begin{figure}[t!]
\begin{centering}
  \includegraphics[width=0.8\linewidth]{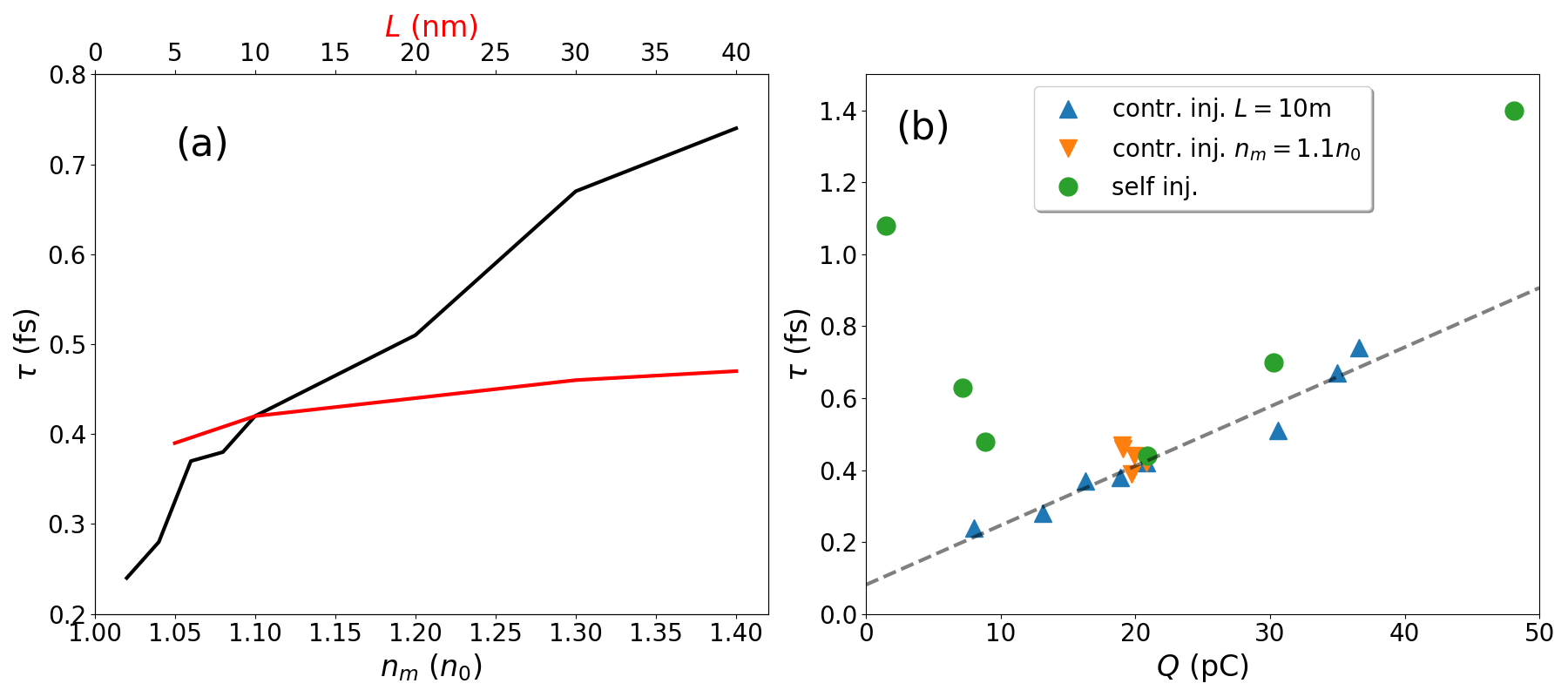}
  \caption{(a) Temporal duration of the injected beam as a function of the maximum peak density $n_m$ for a down-ramp length $L=10~\upmu$m (black line) and as a function of $L$ for $n_m = 1.1 n_0$ (red line). (b) Duration of the injected beam as a function of charge for different data sets: self-injection (green circles),  fixed down-ramp length $L=10~\upmu$m (blue upward triangles) and fixed maximum peak density $n_m = 1.1 n_0$ (orange downward triangles).}
\label{fig:fig3}
\end{centering}
\end{figure}

We now turn to the robustness of the injection process with respect to the charge and duration of the electron beams. Figure~\ref{fig:fig3}(a) shows the effect of a variation of the peak density $n_m$ and the down-ramp length $L$. The beam duration generally increases with $n_m$, with the shortest duration measured for the smallest values of the density peak. A steeper gradient indeed ensures a faster longitudinal expansion of the bubble, which lowers the trapping injection and increases the charge and duration of the injected bunch. Bunches as short as 240~as can be obtained when $n_m=1.02~n_0$, for a 8~pC charge, and all the values of $n_m$ within the tested range yield sub-femtosecond duration. The impact of the down-ramp length is small, with only a $\pm 10$\% variation of the duration when the length is changed between $5~\upmu$m and $40~\upmu$m. This suggests that the exact shape of the density step is not so important, which would facilitate experimental implementation. Note that this conclusion agrees with previous work considering down-ramp injection of more massive electron bunches~\cite{massimo2018numerical}.

We summarize the parameters of the electron beams that were obtained though the different injection methods in Fig.~\ref{fig:fig3}(b) in terms of duration as a function of charge. Electron beams produced from controlled injection generally exhibit a shorter duration for a given charge, which corresponds to a higher current compared with the beams produced from self injection. The behavior for higher charges shows a weak dependence on the injection method, with a steady increase of the duration with the charge. For lower charges though, more pronounced differences appear, as the duration for the self-injection cases reaches a threshold, while very short duration can be reached when the charge is lowered for controlled injection. The duration exhibits a close to linear dependence on the charge in this case. When the injected charge is small and beam loading can be neglected, down-ramp injection lead to a roughly constant electron current. However, in contrast to what is predicted in other studies~\cite{Tooley2017,Ekerfelt2017}, arbitrarily short electron beams cannot be obtained due to the beam elongation caused by the energy spread. Finally, contrary to previous studies with shorter laser duration~\cite{Lifschitz2012,Beaurepaire2014}, we note that CEP effects only lead to very small variations of the electron beam parameters in our simulations -- beam duration only varies by a few percents when tuning the CEP, and can be considered negligible.

\begin{figure}[t!]
  \includegraphics[width=1.0\linewidth]{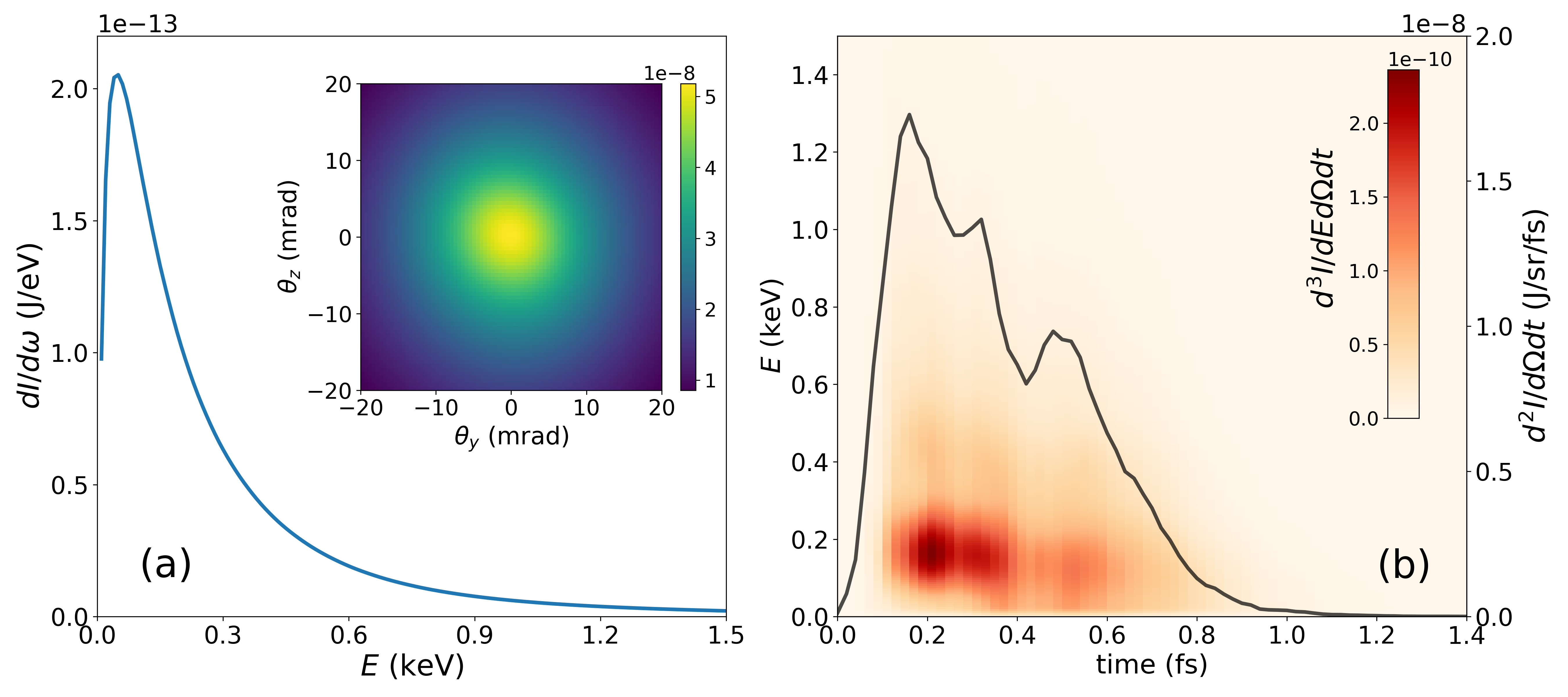}
  \caption{(a) Energy distribution $dI/d\omega$ and angular radiated energy $dI/d\Omega$ (J/sr) for the betatron source with $n_m=1.04~n_0$ and $L=10~\upmu$m. (b) Spectrogram of the radiated energy $d^3I/dEd\Omega dt$ (in J/eV/sr/fs) and temporal profile of the radiation $d^2I/d\Omega dt$(in J/sr/fs, grey line).}
\label{fig:fig4}
\end{figure}
\section{Radiation emission from ultrashort electron bunches}

The radiation emitted by such electron beams is calculated by post-processing the trajectories of the accelerated electrons, using the formula~\cite{Jackson1999}: 
\begin{equation*}
     \frac{d^2I}{d\omega d\Omega} = 
     \frac{e^2}{16\pi^3\epsilon_0c}\left|\int_{-\infty}^{\infty}
    e^{i\omega\left(t-\frac{\mathbf{n}\cdot\boldsymbol{r(t)}}{c}\right)}
    \frac{\mathbf{n}\times\left[\left(\mathbf{n}-\boldsymbol{\beta}(t)\right)\times\dot{\boldsymbol{\beta}}(t)\right]}{(1-\boldsymbol{\beta}(t) \cdot \mathbf{n})^2}dt\right|^2,
\end{equation*}
which gives the radiated energy $d^2I/d\omega d\Omega$ per frequency unit $d\omega$ and per solid angle unit $d\Omega$ in the direction $\mathbf{n}$, and where $\boldsymbol{r}$, $\boldsymbol{\beta}=d\boldsymbol{r}/dt$ and $\dot{\boldsymbol{\beta}}=d\boldsymbol{\beta}/dt$ are the position, speed and acceleration of the particle, respectively, with $e$ the electron charge and $\varepsilon_0$ the vacuum permittivity. Only particles reaching an energy of 5~MeV are taken into account when computing the radiation, as the emission from low-energy particles is negligible.

\begin{figure}[t!]
\begin{centering}
  \includegraphics[width=0.8\linewidth]{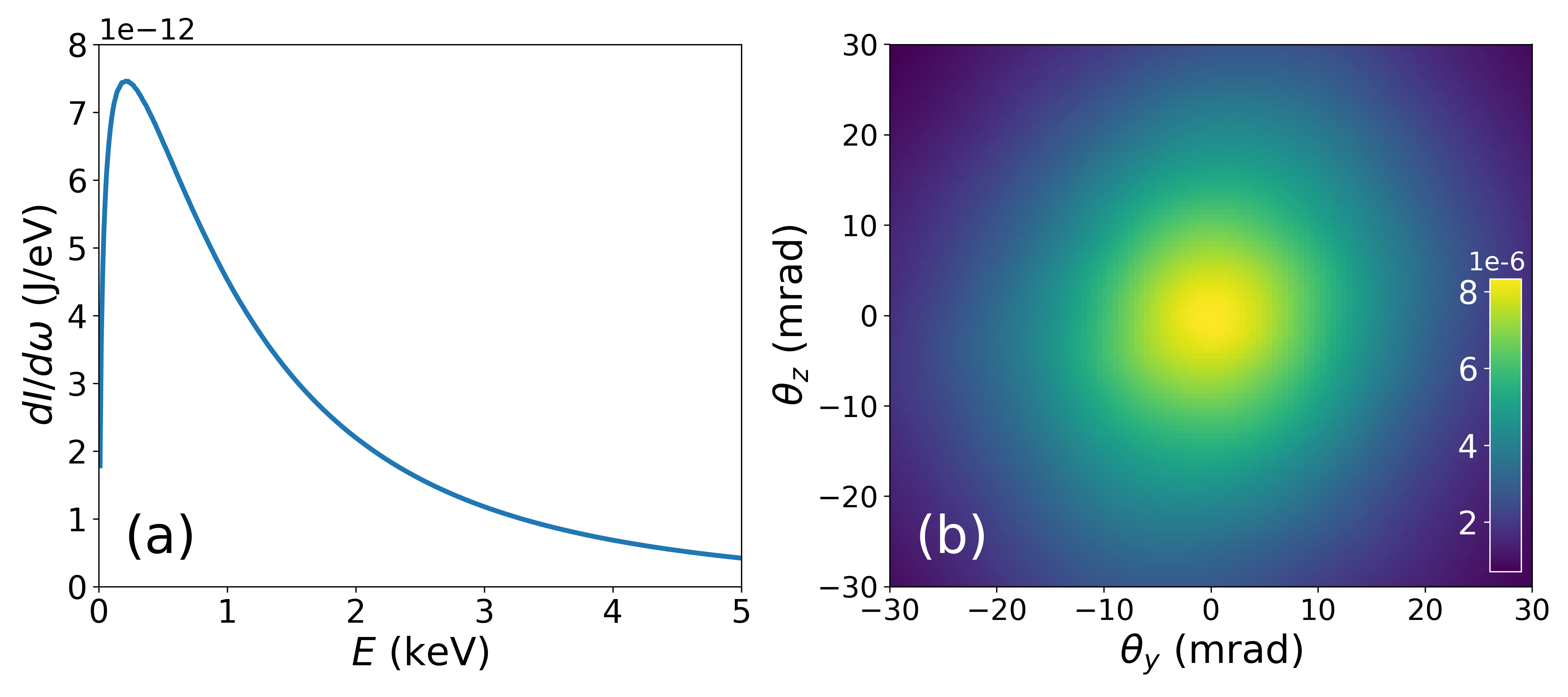}
  \caption{(a) Energy distribution $dI/d\omega$ and (b) angular radiated energy $dI/d\Omega$ (J/sr) for the betatron source corresponding to a constant plateau density of $n_0=4\times10^{19}$~cm$^{-3}$.}
\label{fig:fig5}
\end{centering}
\end{figure}

The energy spectra and angular distribution of the radiation are plotted in Fig.~\ref{fig:fig4}(a) in the case of controlled injection with $n_m=1.04~n_0$ and $L=10~\upmu$m. Due to the relatively low electron energy, the radiation is mainly emitted in the sub-keV domain, with a critical energy $E_c = 200~$eV when fitting the spectrum with $dI/d\omega\propto S(\omega/\omega_c)$, where $S(x)$ is a synchrotron radiation distribution given by $S(x)=x\int_x^{\infty}K_{5/3}(\xi)d\xi$, in which $K_{5/3}$ is a modified Bessel function of the second kind~\cite{Fourmaux2011NJP}. The angle of emission is comparable with standard betatron sources, with divergence of the radiation of 25~mrad (FWHM).
The duration of the betatron radiation is usually inferred from the duration of the electron bunch, but here we compute the spectrogram~\cite{Horny2017}, showing $d^3I/dEd\Omega dt$ on axis in Fig.~\ref{fig:fig4}(b) and its integration over the frequency to get the temporal profile of the radiation on axis in Fig.~\ref{fig:fig4}(c). Most of the radiation is received at early times, with a tail at lower energies arriving at later times, following the structure of the beam. Note that this tail is produced by the lower energy electrons from the first beam and not by the electrons from the second beam. The radiation from the latter is observed with further delay (about 15~fs, not shown here). This yields a duration of 350~as (RMS) for this X-ray pulse. Short duration and small transverse size for the electron beam -- about half a micron in both directions -- due to the high plasma density, lead to a surprisingly high peak brilliance, with $B = 2 \times 10^{22}$~photons per second per mm$^2$ per mrad$^2$ per 0.1\% bandwidth at 100 eV, i.e. similar values to table-top betatron sources obtained with more energetic laser system~\cite{Kneip2010}, albeit at lower photon energies. 
For comparison, the X-ray emitted in the case of self-injection in the matched regime $n_0 = 4\times10^{19}~$cm$^{-3}$ are shown in Fig.~\ref{fig:fig5}. Due to the much higher electron charge and the higher density, this case naturally yields a higher photon energy, as observed in Fig.~\ref{fig:fig5}(a), and the critical energy rises to 1~keV. However, the source duration is also longer, about 5~fs, and the self-injection leads to higher transverse oscillation of the electrons, yielding an increased divergence of $33\times43$~mrad$^2$ (cf.~Fig.~\ref{fig:fig5}). Together with a slightly increased transverse electron beam size ($\sim 1\,\upmu$m), this source yields a more moderate peak brilliance $B = 6 \times 10^{21}$~photons per second per mm$^2$ per mrad$^2$ per 0.1\% bandwidth at 250 eV.

Controlling the injection thus leads to an X-ray source with lower photon energy, but allows for the generation of attosecond X-ray bursts with high brilliance. The main drawback of this source resides in the number of emitted photons being quite low. However, note that the betatron source has not been optimized in this case. This could potentially be improved by several schemes. Using tapered plasma profiles would enable a longer acceleration distance resulting in an increase of both electron energy and emission distance, and prevent the laser defocusing. Schemes to increase the transverse wiggling of the electrons could also boost the betatron emission while in principle not modifying the source duration~\cite{Yu2018, Kozlova2020}. Finally, applying this to structured laser pulses with orbital angular momentum would provide tunable ultra-short X-ray pulses with high photon energy~\cite{Martins2019}.

\section{Conclusion}
Few-cycle laser pulses can be used to generate sub-femtosecond electron bunches under specific conditions. We find, that while operating in the optimal conditions for electron acceleration, bunches produced by self-injection are not particularly short, due to massive injection. Sub-femtosecond duration can be obtained both by operating in the self-injection regime when lowering the density just over the threshold for injection, and by introducing a density down-ramp to control the injection. However, only the latter mechanism yields a short duration for a stable range of parameters and allows for isolated attosecond electron bunches. Finally, the betatron X-ray emission produced by such short electron bunches was characterized, finding attosecond duration, and high brilliance -- despite the relatively low numbers of photons -- owing to the small dimensions of the electron beam. Further optimization of the radiation characteristics produced by this betatron source would open new perspectives for attosecond probing with high repetition rate.

\ack
The authors are grateful for fruitful discussions with O Lundh, L Yi and I Pusztai. This project has received funding from the European Research Council
(ERC) under the European Union's Horizon 2020 research and innovation
programme under grant agreement No 647121, and the Knut och Alice Wallenberg
Foundation. The simulations were performed on resources provided by the Swedish
National Infrastructure for Computing (SNIC) at Chalmers Centre for
Computational Science and Engineering (C$^3$SE) and High Performance
Computing Center North (HPC$^2$N).
\section*{\refname}

\bibliographystyle{iopart-num}

\bibliography{sample.bib} 

\providecommand{\newblock}{}
\begin{thebibliography}{10}
\expandafter\ifx\csname url\endcsname\relax
  \def\url#1{{\tt #1}}\fi
\expandafter\ifx\csname urlprefix\endcsname\relax\def\urlprefix{URL }\fi
\providecommand{\eprint}[2][]{\url{#2}}

\bibitem{Tajima1979}
Tajima T and Dawson J~M 1979 {\em Phys. Rev. Lett.\/} {\bf 43}(4) 267

\bibitem{Mangles_Nature_2004}
Mangles S, Murphy C, Najmudin Z, Thomas A, Collier J, Dangor A, Divall E,
  Foster P, Gallacher J, Hooker C, Jaroszynski D, Langley A, Mori W, Norreys P,
  Tsung F, Viskup R, Walton B and Krushelnick K 2004 {\em Nature (London)\/}
  {\bf 431} 535–538

\bibitem{Geddes_Nature_2004}
Geddes C, Toth C, van Tilborg J, Esarey E, Schroeder C, Bruhwiler D, Nieter C,
  Cary J and Leemans W 2004 {\em Nature (London)\/} {\bf 431} 538

\bibitem{Faure_Nature_2004}
Faure J, Glinec Y, Pukhov A, Kiselev S, S G, Lefebvre E, Rousseau J~P, Burgy F
  and V M 2004 {\em Nature (London)\/} {\bf 431} 541–544

\bibitem{wang2013}
Wang X, Zgadzaj R, Fazel N, Li Z, Yi S~A, Zhang X, Henderson W, Chang Y~Y,
  Korzekwa R, Tsai H~E, Pai C~H, Quevedo H, Dyer G, Gaul E, Martinez M,
  Bernstein A~C, Borger T, Spinks M, Donovan M, Khudik V, Shvets G, Ditmire T
  and Downer M~C 2013 {\em Nature Communications\/} {\bf 4} 1988
  \urlprefix\url{https://doi.org/10.1038/ncomms2988}

\bibitem{Gonsalves2019}
Gonsalves A~J, Nakamura K, Daniels J, Benedetti C, Pieronek C, de~Raadt T~C~H,
  Steinke S, Bin J~H, Bulanov S~S, van Tilborg J, Geddes C~G~R, Schroeder C~B,
  T{\'o}th C, Esarey E, Swanson K, Fan-Chiang L, Bagdasarov G, Bobrova N,
  Gasilov V, Korn G, Sasorov P and Leemans W~P 2019 {\em Phys. Rev. Lett.\/}
  {\bf 122}(8) 084801
  \urlprefix\url{https://link.aps.org/doi/10.1103/PhysRevLett.122.084801}

\bibitem{Bulanov1998}
Bulanov S, Naumova N, Pegoraro F and Sakai J 1998 {\em Phys. Rev. E\/} {\bf
  58}(5) R5257--R5260
  \urlprefix\url{https://link.aps.org/doi/10.1103/PhysRevE.58.R5257}

\bibitem{Geddes2008}
Geddes C~G~R, Nakamura K, Plateau G~R, Toth C, Cormier-Michel E, Esarey E,
  Schroeder C~B, Cary J~R and Leemans W~P 2008 {\em Phys. Rev. Lett.\/} {\bf
  100}(21) 215004
  \urlprefix\url{https://link.aps.org/doi/10.1103/PhysRevLett.100.215004}

\bibitem{Gonsalves2011}
Gonsalves A~J, Nakamura K, Lin C, Panasenko D, Shiraishi S, Sokollik T,
  Benedetti C, Schroeder C~B, Geddes C~G~R, van Tilborg J, Osterhoff J, Esarey
  E, Toth C and Leemans W~P 2011 {\em Nature Physics\/} {\bf 7} 862--866
  \urlprefix\url{https://doi.org/10.1038/nphys2071}

\bibitem{Hansson2015}
Hansson M, Aurand B, Davoine X, Ekerfelt H, Svensson K, Persson A, Wahlstr\"om
  C~G and Lundh O 2015 {\em Phys. Rev. ST Accel. Beams\/} {\bf 18}(7) 071303
  \urlprefix\url{https://link.aps.org/doi/10.1103/PhysRevSTAB.18.071303}

\bibitem{Esarey2002}
Esarey E, Shadwick B~A, Catravas P and Leemans W~P 2002 {\em Phys. Rev. E\/}
  {\bf 65}(5) 056505
  \urlprefix\url{https://link.aps.org/doi/10.1103/PhysRevE.65.056505}

\bibitem{Rousse2004}
Rousse A, Phuoc K~T, Shah R, Pukhov A, Lefebvre E, Malka V, Kiselev S, Burgy F,
  Rousseau J~P, Umstadter D and Hulin D 2004 {\em Phys. Rev. Lett.\/} {\bf
  93}(13) 135005
  \urlprefix\url{https://link.aps.org/doi/10.1103/PhysRevLett.93.135005}

\bibitem{Corde_RevModPhys_2013}
Corde S, Ta~Phuoc K, Lambert G, Fitour R, Malka V, Rousse A, Beck A and
  Lefebvre E 2013 {\em Rev. Mod. Phys.\/} {\bf 85}(1) 1--48
  \urlprefix\url{https://link.aps.org/doi/10.1103/RevModPhys.85.1}

\bibitem{Shah2006}
Shah R~C, Albert F, Ta~Phuoc K, Shevchenko O, Boschetto D, Pukhov A, Kiselev S,
  Burgy F, Rousseau J~P and Rousse A 2006 {\em Phys. Rev. E\/} {\bf 74}(4)
  045401 \urlprefix\url{https://link.aps.org/doi/10.1103/PhysRevE.74.045401}

\bibitem{Kneip2010}
Kneip S, McGuffey C, Martins J, Martins S, Bellei C, Chvykov V, Dollar F,
  Fonseca R, Huntington C, Kalintchenko G, Maksimchuk A, Mangles S, Matsuoka T,
  Nagel S, Palmer C, Schreiber J, Ta~Phuoc K, Thomas A, Yanovsky V, Silva L,
  Krushelnick K and Najmudin Z 2010 {\em Nature Physics\/} {\bf 6} 980--983
  \urlprefix\url{https://doi.org/10.1038/nphys1789}

\bibitem{Kneip2011}
Kneip S, McGuffey C, Dollar F, Bloom M~S, Chvykov V, Kalintchenko G,
  Krushelnick K, Maksimchuk A, Mangles S~P~D, Matsuoka T, Najmudin Z, Palmer
  C~A~J, Schreiber J, Schumaker W, Thomas A~G~R and Yanovsky V 2011 {\em
  Applied Physics Letters\/} {\bf 99} 093701
  \urlprefix\url{https://doi.org/10.1063/1.3627216}

\bibitem{Fourmaux2011}
Fourmaux S, Corde S, Phuoc K~T, Lassonde P, Lebrun G, Payeur S, Martin F,
  Sebban S, Malka V, Rousse A and Kieffer J~C 2011 {\em Opt. Lett.\/} {\bf 36}
  2426--2428 \urlprefix\url{http://ol.osa.org/abstract.cfm?URI=ol-36-13-2426}

\bibitem{Mahieu2018}
Mahieu B, Jourdain N, Ta~Phuoc K, Dorchies F, Goddet J~P, Lifschitz A, Renaudin
  P and Lecherbourg L 2018 {\em Nature Communications\/} {\bf 9} 3276
  \urlprefix\url{https://doi.org/10.1038/s41467-018-05791-4}

\bibitem{Schmid2009}
Schmid K, Veisz L, Tavella F, Benavides S, Tautz R, Herrmann D, Buck A, Hidding
  B, Marcinkevicius A, Schramm U, Geissler M, Meyer-ter Vehn J, Habs D and
  Krausz F 2009 {\em Phys. Rev. Lett.\/} {\bf 102}(12) 124801
  \urlprefix\url{https://link.aps.org/doi/10.1103/PhysRevLett.102.124801}

\bibitem{Guenot2017}
Guénot D, Gustas D, Vernier A, Beaurepaire B, Böhle F, Bocoum M, Lozano M,
  Jullien A, Lopez-Martens R, Lifschitz A and Faure J 2017 {\em Nature
  Physics\/} {\bf 11} 293--296
  \urlprefix\url{https://doi.org/10.1038/nphoton.2017.46}

\bibitem{Lifschitz2012}
Lifschitz A~F and Malka V 2012 {\em New Journal of Physics\/} {\bf 14} 053045
  \urlprefix\url{https://doi.org/10.1088%2F1367-2630%2F14%2F5%2F053045}

\bibitem{Beaurepaire2014}
Beaurepaire B, Lifschitz A and Faure J 2014 {\em New Journal of Physics\/} {\bf
  16} 023023
  \urlprefix\url{https://doi.org/10.1088%2F1367-2630%2F16%2F2%2F023023}

\bibitem{Lundh2011}
Lundh O, Lim J, Rechatin C, Ammoura L, Ben-Ismaïl A, Davoine X, Gallot G,
  Goddet J~P, Lefebvre E, Malka V and Faure J 2011 {\em Nature Physics\/} {\bf
  7} 219--222 \urlprefix\url{https://doi.org/10.1038/nphys1872}

\bibitem{Tooley2017}
Tooley M~P, Ersfeld B, Yoffe S~R, Noble A, Brunetti E, Sheng Z~M, Islam M~R and
  Jaroszynski D~A 2017 {\em Phys. Rev. Lett.\/} {\bf 119}(4) 044801
  \urlprefix\url{https://link.aps.org/doi/10.1103/PhysRevLett.119.044801}

\bibitem{Horny2019}
Horn{\'{y}} V, Petr{\v{z}}{\'{\i}}lka V and Krůs M 2019 {\em Plasma Physics
  and Controlled Fusion\/} {\bf 61} 085018
  \urlprefix\url{https://doi.org/10.1088%2F1361-6587%2Fab2728}

\bibitem{Zhao2019}
Zhao Q, Weng S~M, Chen M, Zeng M, Hidding B, Jaroszynski D~A, Assmann R and
  Sheng Z~M 2019 {\em Plasma Physics and Controlled Fusion\/} {\bf 61} 085015
  \urlprefix\url{https://doi.org/10.1088%2F1361-6587%2Fab249c}

\bibitem{Lifschitz2009}
Lifschitz A, Davoine X, Lefebvre E, Faure J, Rechatin C and Malka V 2009 {\em
  Journal of Computational Physics\/} {\bf 228} 1803 -- 1814 ISSN 0021-9991
  \urlprefix\url{http://www.sciencedirect.com/science/article/pii/S0021999108005950}

\bibitem{Lehe2013}
Lehe R, Lifschitz A, Thaury C, Malka V and Davoine X 2013 {\em Phys. Rev. ST
  Accel. Beams\/} {\bf 16}(2) 021301
  \urlprefix\url{https://link.aps.org/doi/10.1103/PhysRevSTAB.16.021301}

\bibitem{Schmid2010}
Schmid K, Buck A, Sears C~M~S, Mikhailova J~M, Tautz R, Herrmann D, Geissler M,
  Krausz F and Veisz L 2010 {\em Phys. Rev. ST Accel. Beams\/} {\bf 13}(9)
  091301 \urlprefix\url{https://link.aps.org/doi/10.1103/PhysRevSTAB.13.091301}

\bibitem{Dopp2018}
D\"opp A, Thaury C, Guillaume E, Massimo F, Lifschitz A, Andriyash I, Goddet
  J~P, Tazfi A, Ta~Phuoc K and Malka V 2018 {\em Phys. Rev. Lett.\/} {\bf
  121}(7) 074802
  \urlprefix\url{https://link.aps.org/doi/10.1103/PhysRevLett.121.074802}

\bibitem{massimo2018numerical}
Massimo F, Lifschitz A, Thaury C and Malka V 2018 {\em Plasma Physics and
  Controlled Fusion\/} {\bf 60} 034005
  \urlprefix\url{https://doi.org/10.1088%2F1361-6587%2Faaa336}

\bibitem{Ekerfelt2017}
Ekerfelt H, Hansson M, Gallardo~Gonz{\'a}lez I, Davoine X and Lundh O 2017 {\em
  Scientific Reports\/} {\bf 7}
  \urlprefix\url{https://doi.org/10.1038/s41598-017-12560-8}

\bibitem{Jackson1999}
Jackson J~D 1999 {\em Classical electrodynamics\/} 3rd ed (New York, {NY}:
  Wiley) ISBN 9780471309321 \urlprefix\url{http://cdsweb.cern.ch/record/490457}

\bibitem{Fourmaux2011NJP}
Fourmaux S, Corde S, Ta~Phuoc K, Leguay P, Payeur S, Lassonde P, Gnedyuk S,
  Lebrun G, Fourment C, Malka V, Sebban S, Rousse A and Kieffer J 2011 {\em New
  J. Phys.\/} {\bf 13} 033017

\bibitem{Horny2017}
Horný V, Nejdl J, Kozlová M, Krůs M, Boháček K, Petržílka V and Klimo O
  2017 {\em Physics of Plasmas\/} {\bf 24} 063107
  \urlprefix\url{https://doi.org/10.1063/1.4985687}

\bibitem{Yu2018}
Yu C, Liu J, Wang W, Li W, Qi R, Zhang Z, Qin Z, Liu J, Fang M, Feng K, Wu Y,
  Ke L, Chen Y, Wang C, Xu Y, Leng Y, Xia C, Li R and Xu Z 2018 {\em Applied
  Physics Letters\/} {\bf 112} 133503
  \urlprefix\url{https://doi.org/10.1063/1.5019406}

\bibitem{Kozlova2020}
Kozlová M, Andriyash I, Gautier J, Sebban S, Smartsev S, Jourdain N, Chulagain
  U, Azamoum Y, Tafzi A, Goddet J~P, Oubrerie K, Thaury C, Rousse A and
  Ta~Phuoc K 2020 {\em Phys. Rev. X\/} {\bf 10}(1) 011061
  \urlprefix\url{https://link.aps.org/doi/10.1103/PhysRevX.10.011061}

\bibitem{Martins2019}
Luis~Martins J, Viera J, Ferri J and T\"unde F 2019 {\em Scientific Reports\/}
  {\bf 9} \urlprefix\url{https://doi.org/10.1038/s41598-019-45474-8}

\end{thebibliography}

\end{document}